\documentclass[aps,prd,english,superscriptaddress,11pt,notitlepage]{revtex4}
	\usepackage[colorlinks=true, a4paper=true, pdfstartview=FitV,
linkcolor=blue, citecolor=blue, urlcolor=blue]{hyperref}

\usepackage{amsmath}
\usepackage{amssymb}
\usepackage{graphicx}
\usepackage{hhline}
\usepackage{textcomp}
\makeatletter
\usepackage{babel}
\newcommand{\bea}{\begin{eqnarray}}
\newcommand{\eea}{\end{eqnarray}}

\newcommand{\be}{\begin{equation}}
\newcommand{\ee}{\end{equation}}

\usepackage[active]{srcltx}
\begin{document}



\title{Kink-kink solutions in BPS impurity models}

\author{K. S\l awi\'{n}ska}
\affiliation{Institute of Theoretical Physics, Jagiellonian University,
Lojasiewicza 11, Krak\'{o}w, Poland}

\begin{abstract}
We show that BPS-impurity theories may support BPS kink-kink solutions i.e., an energetically degenerated family of solutions describing two kinks at any mutual distance. This requires a singular impurity. As an example we consider the sine-Gordon and $\phi^6$ models coupled with such a BPS impurity. Interestingly, obtained solutions are identical to double sine-Gordon kinks and Christ-Lee kinks respectively. 

We also study the spectral flow on the moduli space. All the modes have an odd number of nodes to cancel the singularity of the impurity. 
\end{abstract}
\maketitle

\vspace*{0.05cm}

\section{Motivation}
Study of Bogomol'nyi-Prasad-Sommerfield (BPS) solitons \cite{B,PS}, that is topologically nontrivial solutions which saturate a pertinent topological bound on energy, is important for various reasons. First of all, the saturation of the bound is equivalent to the fact that the field obeys a lower order differential equation(s) usually called as {\it Bogomol'nyi equation(s)}. Such equations are much simpler than the generic second order Euler-Lagrange equations and therefore, even if not always are analytically solvable, offer a way for an analytical insight into mathematical properties of topological solitons, see e.g., \cite{MS} for a review. 

Secondly, saturation of the bound means that the energy is a linear function of the corresponding topological charge $E=c_0|Q|$ \cite{Fer}. Thus if the Bogomolny equation has a non empty set of solutions in any topological sector then the energy of the soliton of charge $Q=N$ is exactly the same as the energy of the collection of $N$ separated unit charge solitons. This results in the absence of the static force in a multi-soliton state. In other words, individual solitons do not attract nor repeal each other. This happens e.g., for vortices in the Abelian Higgs model at the critical coupling, BPS monopoles and for the self-dual instantons \footnote{There are known models where a Bogomol'nyi equation has nontrivial solutions only in charge one sector, see e.g., a BPS Skyrme model by Harland \cite{H}.}. 

Thirdly, the lack of the static inter-soliton force allows for a semi-analytical understanding of  the dynamics of the BPS solitons. This is because in the lowest order evolution only the zero modes are excited and in a dynamical process the field passes through available, energetically equivalent, BPS solutions. This found an elegant formulation in terms of the geodesic dynamics on the pertinent moduli space \cite{NM-1}. Although the geodesic (zero mode) evolution led to various beautiful results (cf. $\pi/2$ scattering of BPS-vortices \cite{T,S}) it is just the starting point for the understanding of the multi-soliton dynamics. 

It has been recently shown that the massive normal modes, in particular their deformation while we flow on the moduli space, can significantly deform the geodesics dynamics.  In the most extreme case such mode can exist only in the subspace of the moduli space. At the boundary of this subspace the spectral wall phenomenon occurs \cite{SW}. This boundary acts as a barrier for the soliton if it caries some excitation of the mode. Originally this effect has been discovered in (1+1) dimensions in a single scalar field theory coupled with the BPS impurity \cite{BPS-imp-2, BPS-imp-1}. Dynamics of the resulting kink or kink-antikink BPS solutions (both in the presence of the BPS impurity) is clearly affected by the spectral walls \cite{SW, BPS-imp-3}. Very recently, the flow of the spectral structure for the two-vortex  has also been determined. Importantly, one of the modes hits the mass threshold for a certain inter-vortex separation and then disappears for bigger separations \cite{SAL}. As expected this fact gives rise to the spectral wall in the self-dual 2-vortex dynamics \cite{AMRW}.

In generality decreasing (increasing) of the frequency of a massive mode as the BPS solitons approach each other generates the attractive (repulsive) mode induced force \cite{AMMW}. E.g., this leads to chaotic behaviour in the scatterings of excited BPS 2-vortex \cite{R, AMMW}.  

In the current paper we want to present the simplest model which admits a BPS two-soliton solution. Namely, a BPS-impurity model in (1+1) dimensions.  The arising two-kink solution may be viewed as a toy model of two-vortex solution. Therefore studying the flow of its modes as well as its dynamics may help to better understand the time evolution of the BPS two-vortices beyond the geodesic approximation. 

\section{The BPS-impurity models}

We begin with a brief summary of the BPS-impurity framework \cite{BPS-imp-2, BPS-imp-1}. For simplicity, we restrict the presentation to a single scalar field theory in (1+1) dimensions. The main idea is to deform a given theory 
\be
\mathcal{L}=\frac{1}{2}\phi_t^2-\frac{1}{2}\phi_x^2 -  \frac{1}{2}W^2(\phi)
\ee
by an addition of a non-dynamical, background field $\sigma(x)$ in the following way \footnote{We remark that there are many other ways to couple an impurity to a scalar field theory. However, typically they do not lead to a BPS theory. Kinks always are attracted or repealed by the background field. As examples see \cite{imp1, imp2, imp3, imp4}}
\be
\mathcal{L}=\frac{1}{2}\phi_t^2-\frac{1}{2}\left( \phi_x - \sigma W\right)^2 -\phi_xW.
\ee
Here, $U(\phi)=\frac{1}{2}W^2(\phi)$ is the field theoretical potential of the original model, which is assumed to have at least two vacua, while $\phi_t, \phi_x$ denote the derivatives of the field. To see that such a coupling leads to the existence of a Bogomolny equation and, in a consequence, preserves the BPS property in a given topological sector let us consider the energy functional $E$. We find that
\be
E=\int_{-\infty}^\infty \left(  \frac{1}{2}\phi_t^2+\frac{1}{2}\left( \phi_x - \sigma W\right)^2 +\phi_xW\right) dx
\geq  \int_{-\infty}^\infty \phi_xW(\phi) dx =  \int_{\phi(-\infty)}^{\phi(\infty)} W(\phi)d\phi, 
\ee
where the last integral depends only on the model, that is the choice of $W$, and on the boundary condition i.e., on the topology of the solution.

The inequality is saturated if and only if the following impurity modified Bogomolny equations are satisfied
\be
\phi_t=0, \;\;\;  \phi_x - \sigma W(\phi) =0 .
\ee
It is easy to verify that solutions of the Bogomolny equations obey the second order equation of motion
\be
-\phi_{tt}+\phi_{xx} -\sigma_x W-\sigma^2 WW_\phi=0.
\ee
and, therefore, are the static solutions of the impurity deformed model minimizing the energy in the corresponding topological sector. In fact, there is a whole one-parameter family of solutions fulfilling the Bogomolny equation. It is convenient to introduce a new base space variable $y=y(x)$ 
\be
\frac{dy}{dx}=\sigma(x).
\ee
Then the static Bogomolny equation can be brought to the impurity free form
\be
\phi_y(y)-W(\phi)=0
\ee
with solutions $\Phi(y)$ enjoying a translation shift in the new coordinate $y\to y+a$. The continuous parameter $a$, called {\it modulus}, parametrizes the BPS states $\Phi(y(x)+a)$ and give rise to the {\it canonical moduli space} of the energetically equivalent solutions. 

The most important feature of the BPS-impurity framework is that, although the BPS solutions are energetically degenerated, the spectrum of the linear modes varies as we flow on the moduli space i.e., as we change the modulus $a$. This allowed for investigation how the change of the internal modes influences solitonic dynamics leading, for example, to the discovery of the spectral wall phenomenon. We remark that such a spectral structure flow occurs in the case of the vortices in the Abelian Higgs model at the critical coupling. 

The linear mode equation arrises as a small perturbation of the BPS static solution, $\Phi(x;a)+\eta(x,t)$, where the time dependence of the small perturbations is assumed to be of an oscillating form $\eta(x,t)=e^{i\omega t}\eta(x)$. Then we find 
\be
\left. \left( \frac{d^2}{dx^2} - \sigma_x W_\phi - \frac{\sigma^2}{2} \left(W^2\right)_{\phi \phi} \right) \right|_{\phi = \Phi(x;a)} \eta(x;a) = - \omega^2(a) \eta^2(x;a)
\ee
where both the modes $\eta$ as well as the frequencies $\omega$ depend on the position of the modulus space $a$.

In the previous works the impurity has been chosen in a localized form (e.g. $\sigma=1/\cosh^2(x)$) \cite{SW, BPS-imp-2} or in a regular, sign changing form (e.g., $\sigma=\tanh(x)$ \cite{BPS-imp-1, BPS-imp-3}). Such impurities amount to the kink-impurity and the kink-antikink BPS solutions. For example, it is easy to understand the appearance of the antikink-kink BPS solutions. As $x$ goes from $-\infty$ to $+\infty$, the impurity $\sigma=\tanh(x)$ changes from $-1$ to $+1$. Thus, the Bogomolny equation changes from an antikink-type equation ($\phi_x <0$) to a kink-type equation $\phi_x>0$. However, to have a toy model for BPS two-vortices we rather need a kink-kink solution. 

\section{Singular impurity}
In order to construct the family of the BPS two-kink solutions we need two ingredients. First of all, the field theoretical potential $U=W^2$ has to have at least three vacua such that the BPS two-kink solution will pass through all of them. Therefore, as examples we will consider the sine-Gordon model and $\phi^6$ model. Secondly, the impurity has to be chosen in a form which allows to "screen" the usual repulsive kink-kink force. This condition is realized by a singular impurity
\be
\sigma(x)=\coth(x). \label{imp}
\ee
Although the impurity is singular at the origin, the Bogomoly equation can still lead to the regular solutions. However, this implies that the field must tends to the vacuum value at $x=0$. This fact triggers the existence of the two-kink solutions. 
\subsection{The sine-Gordon theory with a BPS impurity}
\begin{figure}
	\includegraphics[height=5cm]{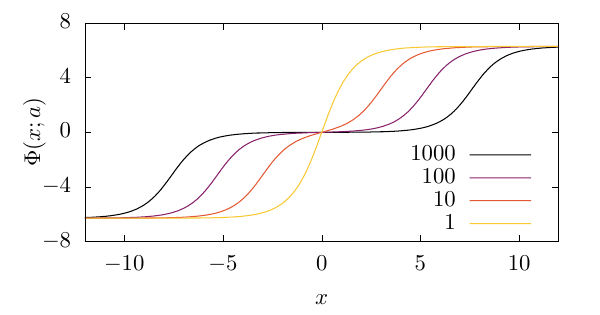} 
	\caption{The BPS solutions for sine-Gordon theory with the BPS-impurity for $a=10^3,10^2,10,1$.}  
	\label{sg_KK}
\end{figure}

We begin with the BPS-impurity version of the famous sine-Gordon model for which $W=2 \sin \frac{\phi}{2}$ (we drop the topological total derivative term)
\begin{equation}
\mathcal{L}_{sG}= \frac{1}{2}\phi_t^2- \frac{1}{2} \left( \phi_x -  2 \sigma \sin \frac{\phi}{2}  \right)^2.
\end{equation}
The corresponding  Bogomolny equation reads
\begin{equation}
\phi_x - 2 \sigma \sin \frac{\phi}{2} =0, \label{bog}
\end{equation}
which for the singular impurity (\ref{imp}) gives the following solutions 
\begin{equation}
\Phi (x; a) = 4\arctan \left( \frac{\sinh x}{a} \right),  \label{BPS-KK}
\end{equation}
where $a \in \mathbb{R}/ \{0\}$ is a continuous parameter, see Fig. \ref{sg_KK}. 

To see that this is a two-soliton solution we first assume that $a=\cosh R$, with $R >0$ ($R<0$ gives the same configurations). Then, after basic algebra we find
\be
\Phi (x;R) = 4\arctan \left( \frac{\sinh x}{\cosh R} \right)=  4\arctan e^{x+R} - 4\arctan e^{R-x},
\ee
which is the sum of the two sine-Gordon kinks located at $\mp R$. As $R\to \infty$ the kinks separate infinitely, while for $R=0$ they are on top of each other. These are, however, not all solutions. The parameter $a$ can be smaller than 1. This can be achieved in the parameterization with modulus $R$ by an assumption that $R$ becomes purely imaginary $R\in [i0,i\pi/2)$. This branch of the solutions describes compressed kinks which finally form a sort of the step-function solutions. To conclude, we found the kink-kink BPS solutions forming a complete moduli space with the modulus $R \in [0,\infty) \cup  [i0,i\pi/2)$ or better $a \in (0,+\infty)$. We remark that for $a<0$ we obtain two-antikink BPS solutions. 

It is interesting that obtained solutions are identical to solutions of the double sine-Gordon model with the potential
\be
U_{d-sG} = \tanh^2 R (1-\cos \phi) +\frac{4}{\cosh^2 R} \left(1+\cos \frac{\phi}{2} \right).
\ee
Of course, in the BPS-impurity deformation of the sine-Gordon, $R$ is the parameter of the {\it solution} while in the double sine-Gordon theory it is the parameter of the {\it potential}. 

\begin{figure}
	\includegraphics[height=4.24cm]{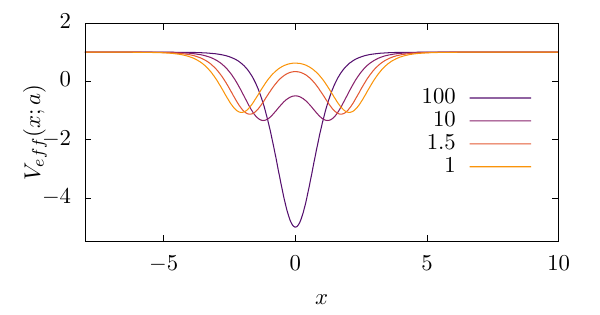} 
		\includegraphics[height=4.24cm]{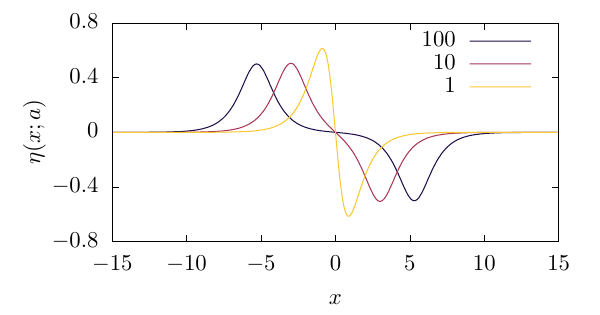} 
	\caption{The potential $V_{eff}(x,a)$ in the spectral problem and the zero-mode for sine-Gordon theory with the BPS-impurity for various $a$ labeled on plots.}  
	\label{F_sg_KK}
\end{figure}

Now, we turn to the normal mode equation, which takes the following form
\be
 \left( \frac{d^2}{dx^2} +\frac{1}{\sinh^2x} \cos \frac{\Phi}{2} - \coth^2(x) \cos \Phi \right) \eta(x;a) = - \omega^2(a) \eta^2(x;a).
\ee
After inserting the BPS two-kink solution we get 
\be
 \left( \frac{d^2}{dx^2} - V_{eff}(x,;a) \right) \eta(x;a) = - \omega^2(a) \eta^2(x;a),
\ee
where
\be
V_{eff}(x;a)=-\frac{1}{\sinh^2x} \frac{a^2-\sinh^2x}{a^2+\sinh^2x} + 2\frac{\cosh^2x}{\sinh^2x} \left[ \left(  \frac{a^2-\sinh^2x}{a^2+\sinh^2x} \right)^2 -\frac{1}{2} \right],
\ee
see Fig. \ref{F_sg_KK}, left panel.
For $a=\cosh R$ this can be viewed as the superposition of the two potential wells generated by each of the sine-Gordon kink. As they approach to each other, that is as $R$ decreases from $+\infty$, the wells merge. When $R$ jumps on imaginary branch the well becomes narrower and its bottom goes to $-\infty$ as $R\to i \frac{\pi}{2}$. We found that the structure of the modes revels a rather nontrivial feature.

First of all, one can easily compute the zero mode related to the existence of the family of the BPS solution (\ref{BPS-KK})
\be
\eta_0(x;a) \sim \partial_a \Phi(x;a)=-\frac{4\sinh(x)}{a^2+\sinh^2(x)}, 
\ee
see Fig. \ref{F_sg_KK}, right panel.
Surprisingly, this mode has one node, located at $x=0$. This suggests that there is a lower energy mode without any node and with $\omega^2<0$. That would be of course the unstable mode. And this is formally the case. E.g., in the limit $a=1$ the effective potential reduces to the Poschl-Teller potential with $\lambda=2$
\be
V_{eff}(x;a=1)=1-\frac{6}{\cosh^2(x)}.
\ee
It has two modes: the previously derived zero mode, $\eta_0(x;1) \sim \sinh(x)/\cosh^2(x)$ and a negative mode, $\eta_{-1}(x) \sim 1/\cosh(x)$ with $\omega^2_{-1}=-3$. Such a negative mode can numerically be found for any $a$. 
However, this negative-like mode seems to be in the contradictory to the fact that the BPS solutions saturate the energy bound. There are no lower energy solution. 

This apparent paradox is resolved if we notice that the "unstable mode" changes the value of the filed at the origin. Indeed, since $\eta_{-1}(x=0;a)\neq 0$ (there is no node), an infinitesimally changed BPS solution, $\phi(x)=\Phi(x;a)+\epsilon \eta_{-1}(x;a)$, deos not have zero value at $x=0$. But such a configuration has infinite energy due to the singular form of the impurity. Hence, the "unstable mode" cannot act on the BPS solution and the zero mode, although possessing one node, is the lowest energy mode hosted by the kink-kink solution.  

One can also say that, as $a \to \infty$, we have asymptotically two zero modes - one on each kink. Their anti-symmetric superposition combines to the zero-mode of the full kink-kink solution. Their symmetric superposition, although potentially with lower energy, cannot not be realized dynamically. 

This rather unusual behaviour is obviously an effect of the specific form of the impurity, which has a singularity at the origin. Thus, all finite energetically solutions must obey a fix-point condition $\phi(x,t)=0$.

Finally, we also comment that, for all $a$, there is no other mode with frequency below the mass threshold. There is only the zero mode. 
\subsection{The $\phi^6$ theory with a BPS impurity}
\begin{figure}
	\includegraphics[height=5cm]{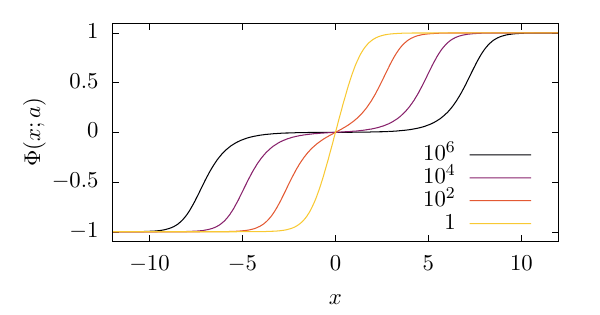} 
	\caption{The BPS solutions for $\phi^6$ with the BPS-impurity for $a=10^6,10^4,10^2,1$.}  
	\label{CL_KK}
\end{figure}

A similar construction can be applied to the $\phi^6$ model, which after coupling to the singular impurity reads
\begin{equation}
\mathcal{L}_{\phi^6}= \frac{1}{2}\phi_t^2- \frac{1}{2} \left( \phi_x -  \sigma \phi (1-\phi^2)  \right)^2.
\end{equation}
Hence now $W=\phi(1-\phi^2)$. 
In this case the potential posses three vacua at $\phi_v=\{-1,0,1\}$.  The BPS kink-kink solutions obey the corresponding Bogomol'nyi equation
\be
 \phi_x -  \sigma \phi (1-\phi^2) =0
\ee
and take the following form
\be
\Phi(x;a)=\frac{ \sinh x}{\sqrt{a +\cosh^2x}}
\ee
with $a \in (-1,\infty)$ being the modulus, see Fig. \ref{CL_KK}. This solutions can be written as the sum of the two $\phi^6$-type kinks: the first interpolating between -1 and 0 vacua and the second one going from 0 to 1 vacuum
\be
\Phi(x;R)= \Phi_{R}( x-R) - \Phi_{R}(-(x+R)),
\ee
where
\be
\Phi_{R}(x)\equiv \frac{1}{\sqrt{\left( 1+e^{-2R}e^{-2x}\right)^2+e^{-2x}}}
\ee
and $R=\ln (2 \sqrt{a})$, for $a>0$. For $a \to \infty$ we get two infinitely separated $\phi^6$ kinks. As $a$ decreases the kinks approach to each other and finally, for $a=0$, they form the usual $\phi^4$ kink. Because this is a regular solution it cannot correspond to the boundary of the moduli space. Indeed, there is the additional branch of the BPS solutions. Once again they can be found by the "complexification" of the position of the kinks $R$. They look as more and more compressed $\phi^4$ kink. 

Interestingly, these solutions exactly coincide with the {\it single}-kink solutions of the Christ-Lee theory where the  field theoretical potential is
\be
U_{CL}=\frac{1}{2\left(a+1 \right)} \left(1+a\phi^2\right)(1-\phi^2)^2.
\ee
Again, in the BPS-impurity deformation of the $\phi^6$ the parameter $a$ labels the solutions while in the Christ-Lee theory it is a parameter of the potential. 

Again we are interested in the flow of the spectral structure on the moduli space i.e, as we change $a$. Now the potential in the eigen-value problem is
\be
V_{eff}(x; a)=-\frac{1}{\sinh^2 x} \left(1- 3 \frac{\sinh^2(x)}{a+\cosh^2(x)} \right) + \mbox{coth}^2(x) \left( 1-\frac{12\sinh^2(x)}{a+\cosh^2(x)} +\frac{15\sinh^4(x)}{(a+\cosh^2(x))^2} \right).
\ee
\begin{figure}
	\includegraphics[height=4.24cm]{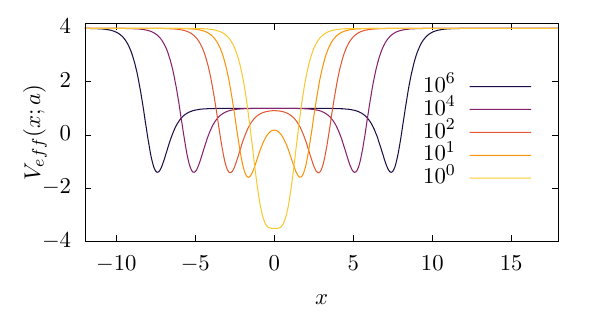} 
		\includegraphics[height=4.24cm]{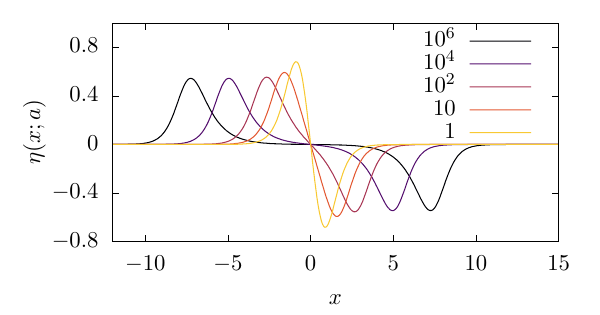} 
	\caption{The potential $V_{eff}(x;a)$ in the spectral problem and the zero-mode  for $\phi^6$ with the BPS-impurity for various $a=10^6,10^4,10^2,0$.}  
	\label{F_CL_KK}
\end{figure}

Once again, only the odd modes are acceptable. The lowest energy is the zero mode with one node at the origin 
\be
\eta_0(x;a) \sim \partial_a \Phi(x;a) = -\frac{1}{2} \frac{\sinh(x)}{\left( a+\cosh^2(x)\right)^{3/2}},
\ee
see Fig. \ref{F_CL_KK}, right panel.

\begin{figure}
	\includegraphics[height=7cm]{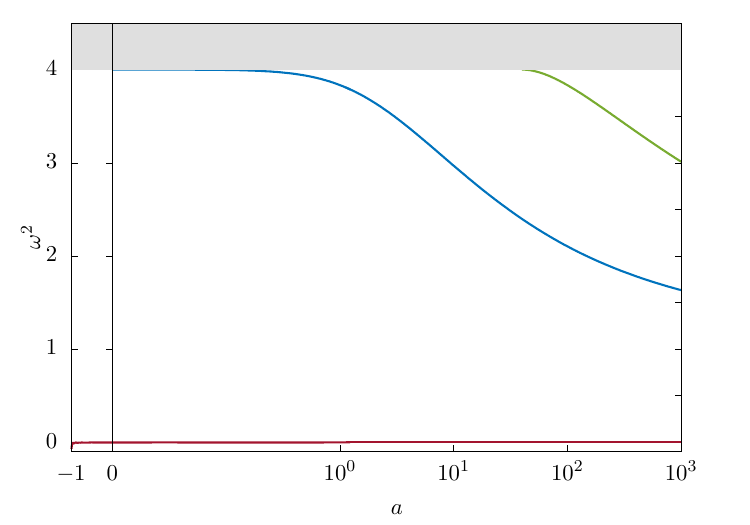} 
	\caption{The flow of the spectral structure of the BPS solutions for the $\phi^6$ with the BPS-impurity as a function of $a$. Lines correspond to discrete odd modes whereas grey region to the continuum spectrum.}  
	\label{cl_modes}
\end{figure}

However, on the contrary to the sine-Gordon type potential, there are massive modes as well. As $a$ grows (to infinity), which corresponds to thee (infinitely) separated kinks, the number of the modes grows to an arbitrary number. This is because the width of the resulting potential well in the eigen-problem grows with $a$ and therefore can support an increasing number of the so-called delocalized modes, see Fig. \ref{F_CL_KK}, left panel. The origin of that is the fact that the mass thresholds in the 0 and $\pm 1$ vacua are different. 

For decreasing $a$ the number of the odd delocalized modes decreases. For $a=0$, $V_{eff}$ reduces to the Poschel-Teller potential with $\lambda=3$
\be
V_{eff}(x; a=0)=4-\frac{12}{\cosh^2(x)}.
\ee
It has therefore only one odd mode, which is our zero mode. All other acceptable bound modes had to hit the mass threshold for larger $a$. In Fig. \ref{cl_modes} we present the flow of the frequencies as a function of $a$. Using Ref. \cite{LY} one may probably find analytical approximation to the modes and values of $a$ at which they cross the mass threshold. 
\section{Summary}

We showed that the BPS-impurity framework can be extended to describe also the BPS kink-kink solutions. Here, the typically repulsive static force between the kinks is completely screened by the impurity and the solitons can be placed at any distance from each other, as expected for the BPS solutions.  

The main modification, in comparison to the previous applications of the BPS-impurity framework, is the form of the impurity. In order to have BPS two-kink solutions it must be singular. Despite of the singularity in the background field, the kinks, the energy density, as well as the spectral problem are well defined. This provides a simple laboratory for studying dynamical problems in higher dimensional BPS models as e.g., Abelian Higgs model at the critical coupling. 

Probably, the most striking features of such BPS theories with a singular impurity is the fact that the zero mode, which is the lowest energy mode, has one node. Due to the finite energy condition the zero-node mode cannot be realized. In fact, because of the same argument, all modes with an even number of nodes are forbidden. Only odd modes can exist.  

Interestingly, both analyzed examples of the families of the BPS solutions describe the well know $90^\circ$ scattering of two solitons, however, realized in (1+1) dimensions. Concretely, we found the BPS solutions where constituent kinks are at a given separation related to the modulus $a$. At some point the  kinks are on top of each other with zero separation. But then the position of the constituents can be continued to purely imaginary values, which in some sense realizes the $90^\circ$ scattering \cite{MORW}. 

Moreover, some of the presented here 2-kink solutions are basically identical to a 2-cycle solutions arising from the iterated $\phi^4$ equation \cite{MOW}. This should be further analyzed. 

Considered here singular impurity, and its application to the construction of the BPS 2-kink solution, can be further investigated in the context of the fat tail kinks 
\cite{BPS-imp-AM}, higher dimensional solitons \cite{BPS-imp-B}. One can also consider quantum 1-loop corrections along the line of \cite{J, BPS-imp-W}. 

We believe that this very simple framework may be helpful in the analysis of the impact of internal modes on the dynamics of the BPS solitons. This includes the role of the mode generated force, the formation of the dynamically induced bound states (due to an attractive mode generated force) or  
 the existence of  the spectral walls. All these phenomena have been very recently reported in the context of the dynamics of the excited vortices in Abelian Higgs models at the critical coupling. 
\section*{Acknowledgements}
KS thanks Tomasz Romanczukiewicz for fruitful discussions.
This work was supported by the Polish National Science Centre (Grant No. NCN 2019/35/B/ST2/00059).


\begin{thebibliography}{99}

\bibitem{B} E. Bogomolnyi, The stability of classical solutions, Sov. J. Nucl. Phys. 24 (1976) 449.

\bibitem{PS} M. Prasad and C. Sommerfield, An Exact Classical Solution for the t Hooft Monopole and the Julia-Zee Dyon, Phys. Rev. Lett. 35 (1975) 760.

\bibitem{MS} N. Manton and P. Sutcliffe, Topological Solitons, Cambridge University Press, Cambridge U.K., 2004.

\bibitem{Fer} C. Adam, L. A. Ferreira, E. da Hora, A. Wereszczynski, W. J. Zakrzewski, Some aspects of self-duality and generalised BPS theories, JHEP 1308 (2013) 062. 

\bibitem{H} D. Harland, Topological energy bounds for the Skyrme and Faddeev models with massive pions, Phys. Lett. B 728 (2014) 518.

\bibitem{NM-1} N. S. Manton, A remark on the scattering of BPS monopoles, Phys. Lett. 110B, 54 (1982).

\bibitem{T} C. H. Taubes, Arbitrary N-vortex solutions to the first order Ginzburg-Landau equations, Commun. Math. Phys. 72, 277 (1980).

\bibitem{S} T. M. Samols, Vortex scattering, Commun. Math. Phys. 145, 149 (1992).

\bibitem{SW} C. Adam, K. Oles, T. Romanczukiewicz, and A. Wereszczynski, Spectral Walls in Soliton Collisions, Phys. Rev. Lett. 122, 241601 (2019).

\bibitem{BPS-imp-2} C. Adam, T. Romanczukiewicz, A. Wereszczynski, The $\phi^4$ model with the BPS preserving defect, JHEP 1903 (2019) 131.

\bibitem{BPS-imp-1} C. Adam, K. Oles, J. M. Queiruga, T. Romanczukiewicz and A. Wereszczynski, Solvable self-dual impurity models, JHEP 1907 (2019) 150.


\bibitem{imp1} B.A. Malomed, Perturbative analysis of the interaction of a $\phi^4$ kink with
inhomogeneities, J. Phys. A 25 (1992) 755.

\bibitem{imp2}  B. Piette and W.J. Zakrzewski, Scattering of sine-Gordon kinks on potential wells, J. Phys. A 40 (2007) 5995. 

\bibitem{imp3} Z. Fei, Y. S. Kivshar, and L. Vazquez, Resonant kink-impurity interactions in the sine-Gordon model, Phys. Rev. A 45, 6019 (1992).

\bibitem{imp4} Z. Fei, L. Vazquez and Y.S. Kivshar, Resonant kink impurity interactions in the $\phi^4$
model, Phys. Rev. A 46 (1992) 5214. 


\bibitem{BPS-imp-3} C. Adam, K. Oles, T. Romanczukiewicz and A. Wereszczynski, Kink-antikink scattering in the $\phi^4$ model without static intersoliton forces, Phys. Rev. D 101 (2020) 105021.

\bibitem{SAL}  A. Alonso Izquierdo, W. Garcia Fuertes, N. S. Manton and J. Mateos Guilarte, Spectral Flow of Vortex Shape Modes over the BPS 2-Vortex Moduli Space,  J. High Energy Phys.  01 (2024) 020.

\bibitem{AMRW} A. Alonso Izquierdo, J. Mateos Guilarte, M. Rees, A. Wereszczynski, Spectral wall in collisions of excited Abelian Higgs vortices, 2406.05725. 

\bibitem{AMMW} A. Alonso Izquierdo, N. S. Manton, 
  J. Mateos Guilarte, A. Wereszczynski, Collective Coordinate Models for 2-Vortex Shape Mode Dynamics, arXiv:2405.20757.  

\bibitem{R} S. Krusch, M. Rees, T. Wyniard,  {\it Scattering of Vortices with Excited Normal Modes}, arXiv:2406.04164.  

\bibitem{CL} N. H. Christ and T. D. Lee, Quantum Expansion of Soliton Solutions, Phys. Rev. D 12 (1975) 1606.

\bibitem{MORW} N. S. Manton, K. Oles, T. Romanczukiewicz, and A. Wereszczynski, Kink moduli spaces: Collective coordinates reconsidered, Phys. Rev. D 103 (2021)  025024.

\bibitem{MOW} N. S. Manton, K. Oles, and A. Wereszczynski, Iterated $\phi^4$ kinks, JHEP 10 (2019) 086. 

\bibitem{BPS-imp-AM} J. G. F. Campos, A. Mohammadi, Collisions between kinks with long-range tails: a simple and efficient method, JHEP 02 (2024) 056. 

\bibitem{BPS-imp-B} D. Bazeia, M. A. Liao, M. A. Marques, Impurity-doped scalar fields in arbitrary dimensions, Phys. Lett. B 846 (2023) 138262. 

\bibitem{J} J. Evslin, C. Halcrow, T. Romanczukiewicz, A. Wereszczynski, Spectral walls at
one loop, Phys. Rev. D 105 (2022) 125002.

\bibitem{BPS-imp-W} I. Takyi, H. Weigel, Quantum effects of solitons in the self-dual impurity model, Phys. Rev. D 107 (2023) 036003.

\bibitem{LY} L. Long, Y. Jiang, Solving the Spectral Problem via the Periodic Boundary Approximation in $\phi^6$ Theory, 2404.13310. 


\end{thebibliography}
\end{document}